\documentclass[11pt]{tibop-article}
\TIBauthor[Penell and Dailisan]{Juri Penell\affOne\orcidlink{0009-0002-3110-0429}\lastand
            Damian Dailisan\affOne\orcidlink{0000-0003-2899-9082}
          }
\TIBaffiliations{\affOne Computational Social Science ETH Zurich, Switzerland\medskip\\
	*Correspondence: Juri Penell, jpenell@ethz.ch
	
}

\TIBtitle{GROSS: German Rail Open-Source SUMO Scenario}
\TIBsubtitle{}
\TIBbundlename[SUMO]{SUMO Conference}
\TIBpublisheddate{2026-06-01}
\TIBabstract{
Microscopic simulation enables reproducible evaluation in intelligent transportation systems, yet most open SUMO scenarios and toolchains remain road-traffic centric, leaving rail underrepresented despite its importance for public transport and its sensitivity to network-wide disruptions. We present the German Rail Open-Source Scenario (GROSS), an open pipeline that combines OpenStreetMap railway infrastructure with GTFS schedules to generate nation-scale rail scenarios for SUMO (Simulation of Urban MObility). Existing conversions often rely on geometry-only stop-to-track matching and inconsistent platform/track assignments, which can create routing anomalies and unstable simulations dominated by teleportation artefacts. GROSS addresses this with topology-aware stop mapping via a hierarchical station model, followed by station-level routing with validation and targeted repair. Across multiple German regions, GROSS reduces average teleportations per vehicle by a factor of 1.7--76.8$\times$, shortens delays compared to the vanilla SUMO pipeline, and it enables end-to-end generation of a Germany-wide scenario with 35\,925 trips for comparisons with operator-reported delay statistics. While the remaining long delays highlight limitations in available timetable metadata and rail dispatch modeling, GROSS lowers the barrier to building scalable, fully open rail simulations and to studying delay propagation at country scale.
}
\TIBkeywords{Rail simulation, Open-source, Routing}
\usepackage{xfrac}
\usepackage{booktabs}
\usepackage{tabularx}
\usepackage{siunitx}
\usepackage{graphicx} % for \scalebox
\usepackage{tikz}
\usetikzlibrary{positioning,calc,fit,backgrounds,matrix,arrows.meta}
\usepackage{amssymb}
\usepackage{printlen}
\usepackage{algorithm}
\usepackage[noend]{algpseudocode}
\uselengthunit{in}
\usepackage{subcaption}
\usepackage{chngcntr} % reset figure/table numbering within appendix sections
\begin{document}
\maketitle

\section{Introduction}
Microscopic traffic simulation plays a central role in intelligent transportation systems (ITS) research, enabling reproducible evaluation of control strategies, infrastructure design, and mobility policies under realistic operating conditions \cite{lopez2018Microscopic}.
Among existing simulation platforms, SUMO is widely adopted due to its open-source nature and ability to support large-scale, network-level experiments.
Consequently, substantial effort has been devoted to the development of open traffic scenarios that approximate real-world mobility patterns \cite{Bieker2015Bologna, Codeca2017LuST, Codeca2017MoST, Schrab2023BeST, Yamazaki2023ToST, argota2021barcelonatwin}.
Such traffic scenarios are also widely used as a controllable testbed for data-driven, learning-based traffic research~\cite{sumorl,Korecki2023Disruptions,Korecki2024Democratizing,Dubey2025}.

To date, however, most publicly available large-scale scenarios are predominantly road-traffic oriented. Well-established examples for various cities, such as the Luxembourg \cite{Codeca2017LuST}, Monaco \cite{Codeca2017MoST}, Tokyo \cite{Yamazaki2023ToST}, and Berlin \cite{Schrab2023BeST} SUMO traffic scenarios focus primarily on private vehicular mobility, typically relying on demographic data, traffic counts, or calibrated demand generation pipelines. These scenarios have significantly improved reproducibility and comparability in ITS research, yet they capture only a limited subset of modern mobility systems.

In contrast, rail-based transportation constitutes a fundamental component of mobility in many European countries, particularly in Germany, where regional, suburban, and long-distance rail services form the backbone of daily commuting and intercity travel \cite{buehler2017Reducing,heuermann2019effect,boehm2021potential}.
Despite its importance, rail mobility remains underrepresented in open, reusable simulation scenarios. This gap is notable given that railway systems are highly susceptible to network-wide disruptions: infrastructure failures, capacity bottlenecks, and tightly coupled timetables can lead to cascading delays that propagate across large geographic areas \cite{daniotti2024Systemic, Cats2016vulnerability}. Such phenomena significantly degrade service reliability and hamper passenger mobility, motivating increasing interest in robustness analysis and disruption management \cite{dewilde2011Defining}.

Simulation-based approaches are particularly well-suited for studying these effects, as they allow controlled experimentation under repeatable conditions. However, the lack of openly available, large-scale rail scenarios has limited systematic investigation of railway operations within the ITS community. At the same time, the availability and quality of open-access mobility data have improved substantially. OpenStreetMap provides detailed representations of railway infrastructure \cite{osmfoundation}, while the General Transit Feed Specification (GTFS) offers a standardized description of schedules, stops, and services \cite{delfi2025index}. Together, these data sources enable the construction of realistic rail simulation environments without reliance on proprietary datasets.

In this work, we present a pipeline for generating country-scale rail simulation scenarios in SUMO \cite{lopez2018Microscopic,alvarez_lopez_2026_18406080} using open-access OSM and GTFS data.
We apply our pipeline to create train-focused simulation scenarios in Germany.
Unlike existing SUMO pipelines and baseline examples, we show that our approach improves operational realism by reducing the amount of deadlocks and artificial teleportations that occur at junctions in the simulation.
We then demonstrate that our pipeline can be effectively scaled to the national level and create a Germany wide rail scenario.
We characterize the level of realism these scenarios can achieve by comparing them to existing real-world rail statistics from the Deutsche Bahn (DB).
Our rail scenario can serve as a complement to other existing car-centric scenarios, leading to a more complete representation of mobility systems and providing a foundation for future studies on railway operations, delay propagation, and multimodal transportation analysis \cite{litmanMore}.

\section{Related Work}
\subsection{Open SUMO Traffic and Rail Scenarios}
The SUMO community has released a number of widely used open scenarios to foster reproducible experimentation with microscopic traffic simulation \cite{lopez2018Microscopic}. Early work, such as the Bologna scenario, demonstrated end-to-end scenario creation from open data and helped establish common practices for network extraction, demand generation, and calibration \cite{Bieker2015Bologna}. More recent city-scale scenarios---including LuST (Luxembourg) \cite{Codeca2017LuST,Codeca2017MoST,Yamazaki2023ToST,Schrab2023BeST}, MoST (Monaco) \cite{Codeca2017MoST}, ToST (Tokyo) \cite{Yamazaki2023ToST}, and BeST (Berlin) \cite{Schrab2023BeST}---provide curated networks and demand models that are frequently used as baselines and benchmarks. However, these scenarios primarily focus on road traffic; rail infrastructure and train operations are either not included or play only a minor role.

Compared to road-traffic scenarios, openly available large-scale rail scenarios for SUMO remain scarce, despite SUMO's support for rail vehicle dynamics and signaling abstractions. This limits systematic investigation of railway-specific phenomena such as delay propagation, conflicts at junctions, and the interplay between timetables and infrastructure capacity.

\subsection{Timetable Robustness and Delay Propagation}
Robustness analysis and disruption management have long been studied in railway operations research. Ref.~\cite{dewilde2011Defining} formalizes notions of timetable robustness and discusses how slack and buffer times relate to delay recovery. At the network level, partial capacity degradations can lead to disproportionate impacts on passenger service, motivating vulnerability- and resilience-oriented evaluation methodologies \cite{Cats2016vulnerability}. Recent work further emphasizes the systemic nature of delay cascades and proposes models and mitigation strategies that account for network-wide coupling effects \cite{daniotti2024Systemic}. Our work complements this literature by providing an openly reproducible, country-scale simulation scenario that enables controlled experiments on delay propagation under realistic operating constraints.

\subsection{Scenario Construction Pipelines and Practical Limitations}
Many SUMO tutorials and scenario-generation pipelines are designed with road traffic in mind, and their rail-related examples are typically small, synthetic, or hand-crafted. In large rail networks, naive conversion of junction topology and right-of-way rules can lead to deadlocks, unrealistic blocking, and excessive teleportations, reducing the validity of operational studies. By focusing on rail-specific network processing and schedule-driven operations, our pipeline aims to bridge the gap between tutorial-scale examples and country-scale railway simulations.

\section{Methodology}
\newcommand{\code}[1]{\texttt{\detokenize{#1}}}

\begin{figure}[tb]
    \centering
    \small
    \begin{tikzpicture}[
        boxmat/.style={
            matrix of nodes,
            draw=black,
            fill=white,
            inner sep=4pt,
            nodes={anchor=west, align=left},
            row sep=1pt,
            column sep=3pt
        },
        mainarrow/.style={->, line width=0.8pt},
        % subpoint arrow: end exactly on the pill edge (no shortening)
        subturn/.style={->, >= {Latex[length=4pt,width=3pt]}, line width=0.6pt},
        % per-subpoint "pill" (shifted right; extra left padding; white text on green)
        subpill/.style={
            draw=green!60!black,
            fill=green!60!black,
            text=white,
            line width=0.6pt,
            rounded corners=3pt,
            inner ysep=1pt,
            inner xsep=4pt,
            xshift=6pt
        }
    ]
        % All boxes are matrices (so we can structure “subpoints” inside them)
        \matrix (OSM) [boxmat] at (0,1) { OSM\\ };
    
        \matrix (netconvert) [boxmat,
            column sep=2pt,
            column 1/.style={nodes={minimum width=5pt, inner sep=0pt}},
            column 2/.style={nodes={inner sep=0pt}},
            row 2/.style={nodes={inner sep=1pt}},
            row 2 column 2/.style={nodes=subpill}
        ] at (3,1) {
            {} & {\hspace*{-4pt}Network}\\[2pt]
            {} & {\scriptsize netconvert}\\
        };
    
        \matrix (GTFS) [boxmat] at (0,-3.8) { GTFS\\ };
        \matrix (Stations) [boxmat] at (3,-2) { Stations\\ };
    
        \matrix (inRou) [boxmat,
            column sep=2pt,
            column 1/.style={nodes={minimum width=5pt, inner sep=0pt}},
            column 2/.style={nodes={inner sep=0pt}},
            row 2/.style={nodes={inner sep=1pt}},
            row 2 column 2/.style={nodes=subpill}
        ] at (6,-0.8) {
            {} & {\hspace*{-4pt}Inner Router}\\[2pt]
            {} & {\scriptsize duarouter}\\
        };
    
        \matrix (outRou) [boxmat] at (9.5,-2) { Outer Router\\ };
    
        \matrix (sim) [boxmat,
            column sep=2pt,
            column 1/.style={nodes={minimum width=5pt, inner sep=0pt}},
            column 2/.style={nodes={inner sep=0pt}},
            row 2/.style={nodes={inner sep=1pt}},
            row 2 column 2/.style={nodes=subpill}
        ] at (13,-0.5) {
            {} & {\hspace*{-4pt}Simulation}\\[2pt]
            {} & {\scriptsize SUMO}\\
        };
    
        \draw[subturn]
          let \p1 = (netconvert-2-1.north),
              \p2 = (netconvert-2-2.west),
              \p3 = ([xshift=-4pt]netconvert-2-2.west),
              \n1 = {\x3-\x1}, \n2 = {2pt} in
          (\x1+\n2,\y3+\n1-\n2) arc[start angle=180,end angle=270,radius=\n1-\n2] -- (\x2,\y2);
    
        \draw[subturn]
          let \p1 = (inRou-2-1.north),
              \p2 = (inRou-2-2.west),
              \p3 = ([xshift=-4pt]inRou-2-2.west),
              \n1 = {\x3-\x1}, \n2 = {2pt} in
          (\x1+\n2,\y3+\n1-\n2) arc[start angle=180,end angle=270,radius=\n1-\n2] -- (\x2,\y2);
    
        \draw[subturn]
          let \p1 = (sim-2-1.north),
              \p2 = (sim-2-2.west),
              \p3 = ([xshift=-4pt]sim-2-2.west),
              \n1 = {\x3-\x1}, \n2 = {2pt} in
          (\x1+\n2,\y3+\n1-\n2) arc[start angle=180,end angle=270,radius=\n1-\n2] -- (\x2,\y2);
    
        % Group background (stations + routers)
        \begin{scope}[on background layer]
            % darker / less transparent blue backdrop
            \node[fill=blue!25, draw=none, rounded corners,
                  fit=(Stations) (inRou) (outRou), inner sep=10pt] {};
        \end{scope}
    
        \draw[mainarrow] (OSM) -- (netconvert);
    
        % If multiple arrows enter from the same side, separate them by 4pt (±2pt).
        \draw[mainarrow,rounded corners=16pt] (OSM)  |- ([yshift= 4pt]Stations.west);
        \draw[mainarrow,rounded corners=16pt] (GTFS) |- ([yshift=-4pt]Stations.west);
    
        \draw[mainarrow,rounded corners=16pt] (Stations)  |- ([yshift=-4pt]inRou.west);
        \draw[mainarrow,rounded corners=16pt] (netconvert) |- ([yshift= 4pt]inRou.west);
        % ragged-corner connectors (horizontal then vertical) with rounded corners r=16pt
        \draw[mainarrow,rounded corners=16pt] (GTFS) -| (outRou);
        \draw[mainarrow] (Stations) -- (outRou);
        \draw[mainarrow,rounded corners=16pt] (inRou) -| (outRou);
        \draw[mainarrow,rounded corners=16pt] (outRou) -| (sim);
        \draw[mainarrow,rounded corners=16pt] (netconvert) -| (sim);
    \end{tikzpicture}
    \caption{Flowchart of GROSS pipeline architecture. Each box represents a discrete processing step, with green highlights identifying the use of a SUMO tool within a step. The blue-shaded region signifies the public transport processing block, which is comparable to \textnormal{\texttt{gtfs2pt}}.}
    \label{fig:flowchart}
\end{figure}
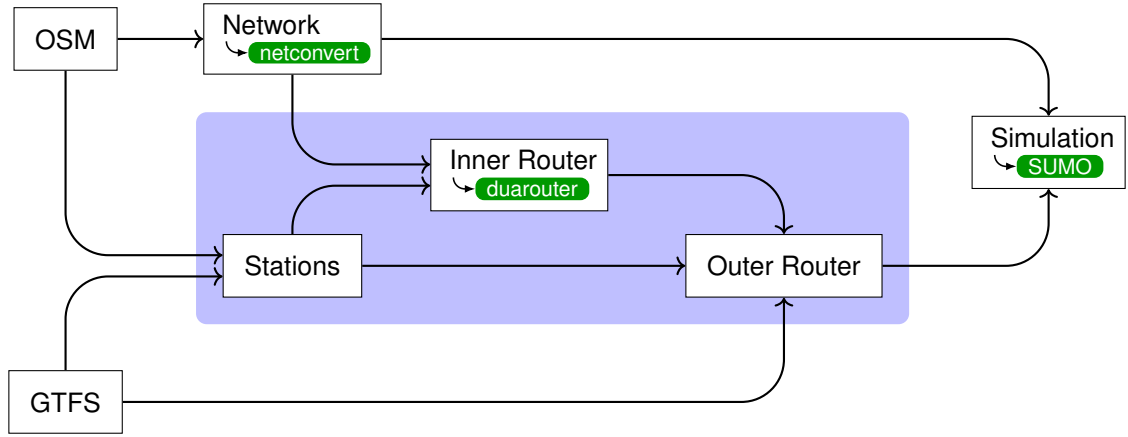

In this section, we describe the methodology and pipeline used to create GROSS (Fig.~\ref{fig:flowchart}). 
In the following subsections, we will describe the following:
(\ref{subsec:data_acquisition})~\nameref{subsec:data_acquisition}, 
(\ref{subsec:baseline})~\nameref{subsec:baseline}, 
(\ref{subsec:network_creation})~\nameref{subsec:network_creation}, 
(\ref{subsec:station_analysis})~\nameref{subsec:station_analysis}, 
(\ref{subsec:routing})~\nameref{subsec:routing}, and 
(\ref{subsec:sim_experiments})~\nameref{subsec:sim_experiments}.

\subsection{Data}
\label{subsec:data_acquisition}
The pipeline relies on two complementary open data sources that together provide both the physical infrastructure and the operational schedule of a rail network: OpenStreetMap (OSM) provides the transport infrastructure \cite{osmfoundation}, and the General Transit Feed Specification (GTFS) provides the passenger train schedules.  

The OSM dataset allows us, after intensive processing, to extract information on train tracks, signals, station platforms, stop positions, and other physical infrastructure. The GTFS records contain the routes, timetables, and basic stop information. Specifically, we use GTFS feeds provided by Delfi~\cite{delfi2025index}, which also include additional metadata (in the form of \texttt{IFOPT} numbers) that help cross-reference the GTFS stops to the OSM stations. Although this additional metadata helped improve train routing in this work, the pipeline is designed to still function if explicit cross-referencing between GTFS and OSM is absent.

\subsection{Baseline}
\label{subsec:baseline}
We benchmark our processing pipeline against the typical workflow one would use following the standard SUMO tools (v1.26), as outlined in the sumo-berlin repository of SUMO~\cite{DLR_TS_sumo_berlin}. Given an OSM network and GTFS schedules, this baseline approach primarily uses \texttt{netconvert} to create the rail network with candidate stops, and \texttt{gtfs2pt.py} to map the GTFS routes onto the network. We then patch the output of \texttt{gtfs2pt} to remove trips that were filtered out due to incorrect tagging (e.g. buses and trams incorrectly tagged as trains), and split trips at segments that would require impossible speeds (Tab. \ref{tab:repair}).

\begin{figure}[tb]
\centering
\begin{subfigure}[t]{0.35\textwidth}
    \begin{tikzpicture}[scale=0.45]

      % Outer quarter circle split into segments
      \draw[line width=2pt, draw=blue] (11,0) -- (11,1);
      \draw[line width=2pt] (11,1) -- (11,3);
      \draw[line width=2pt] (11,3) arc[start angle=0,end angle=90,radius=8];
      \draw[line width=2pt] (3,11) -- (1,11);
      \draw[line width=2pt, draw=blue] (1,11) -- (0,11);
      
      % Connection lines
      \draw[line width=2pt, draw=blue] (11,1) -- (10,3);
      \draw[line width=2pt, draw=blue] (3,10) -- (1,11);

      % Inner quarter circle split into segments
      \draw[line width=2pt, draw=orange!70] (10,0) -- (10,3);
      \draw[line width=2pt, draw=red] (10,3) arc[start angle=0,end angle=90,radius=7];
      \draw[line width=2pt, draw=orange!70] (3,10) -- (0,10);

      % Arrows parallel but shifted
      \draw[->, line width=1pt, draw=blue] (11.3,0.2) -- (11.3,0.8);
      \draw[->, line width=1pt, draw=blue] (0.8,11.3) -- (0.2,11.3);
    
      \draw[->, line width=1pt, draw=orange!70, shift={(-0.3,0)}] (10,2) -- (10,1);
    
        \draw[->, line width=1pt, draw=blue] 
        (3+7.3*cos 37.5, 3+7.3*sin 37.5) arc[start angle=37.5,end angle=52.5,radius=7.3];
    
      % Inner arc 15-degree arrow
      \draw[->, line width=1pt, draw=orange!70] 
        (3+6.7*sin 37.5, 3+6.7*cos 37.5) arc[start angle=52.5,end angle=37.5,radius=6.7];
      
      \draw[->, line width=1pt, draw=orange!70, shift={(0,-0.3)}] (1,10) -- (2,10);
    
      \draw[->, line width=1pt, draw=blue, shift={(0.2,0.2)}] (10.5,2) -- (10.25,2.5);
      \draw[->, line width=1pt, draw=blue, shift={(0.2,0.2)}] (2.5,10.25) -- (2,10.5);
    \end{tikzpicture}
    \caption{Aggressive Routing}
    \label{fig:aggressiveRouting}

\end{subfigure}
\begin{subfigure}[t]{0.35\textwidth}
    \centering
    \begin{tikzpicture}[scale=0.45]

      % Outer quarter circle split into segments
      \draw[line width=2pt, draw=blue] (11,0) -- (11,1);
      \draw[line width=2pt, draw=blue] (11,1) -- (11,3);
      \draw[line width=2pt, draw=blue] (11,3) arc[start angle=0,end angle=90,radius=8];
      \draw[line width=2pt, draw=blue] (3,11) -- (1,11);
      \draw[line width=2pt, draw=blue] (1,11) -- (0,11);
      
      % Connection lines
      \draw[line width=2pt] (11,1) -- (10,3);
      \draw[line width=2pt] (3,10) -- (1,11);

      % Inner quarter circle split into segments
      \draw[line width=2pt, draw=orange!70] (10,0) -- (10,3);
      \draw[line width=2pt, draw=orange!70] (10,3) arc[start angle=0,end angle=90,radius=7];
      \draw[line width=2pt, draw=orange!70] (3,10) -- (0,10);

      % Arrows parallel but shifted
      \draw[->, line width=1pt, draw=blue, shift={(0.3,0)}] (11,1) -- (11,2);
      \draw[->, line width=1pt, draw=blue, shift={(0,0.3)}] (2,11) -- (1,11);

        \draw[->, line width=1pt, draw=blue] 
        (3+8.3*cos 37.5, 3+8.3*sin 37.5) arc[start angle=37.5,end angle=52.5,radius=8.3];
    
      % Inner arc 15-degree arrow
      \draw[->, line width=1pt, draw=orange!70] 
        (3+6.7*sin 37.5, 3+6.7*cos 37.5) arc[start angle=52.5,end angle=37.5,radius=6.7];

      \draw[->, line width=1pt, draw=orange!70, shift={(-0.3,0)}] (10,2) -- (10,1);
      \draw[->, line width=1pt, draw=orange!70, shift={(0,-0.3)}] (1,10) -- (2,10);

      %\draw[->, line width=1pt, draw=blue, shift={(0.3,0)}] (11,2) -- (11,1);
      %\draw[->, line width=1pt, draw=blue, shift={(0,0.3)}] (1,11) -- (2,11);
    
      %\draw[->, line width=1pt, draw=blue] (11.3,2) -- (11.3,2.5);
      %\draw[->, line width=1pt, draw=blue] (2.5,11.3) -- (2,11.3);
    \end{tikzpicture}
    \caption{Desired Routing}
    \label{fig:desiredRouting}

\end{subfigure}
\caption{Curve Routing with clockwise route (orange), counter-clockwise route (blue), and both routes conflicting (red)}
\label{fig:curveRouting}

\end{figure}
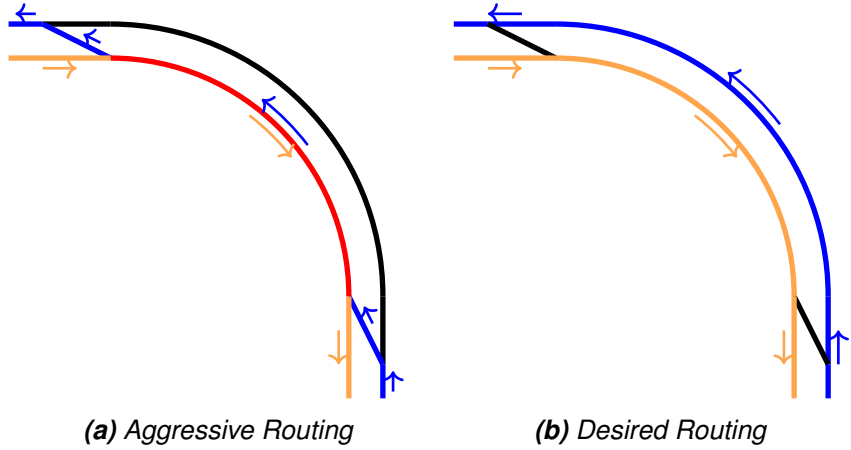
\subsection{Network}
\label{subsec:network_creation}
The rail network used for the simulation is generated using the SUMO tool \texttt{netconvert}. We use it to apply changes and convert the OSM data into a simulation network where all tracks are modeled as bidirectional edges because directional edges introduce non-recoverable anomalies. Stop positions, representing the exact location a train comes to a halt at a station, are taken as the simulation stop locations. We penalize crossovers to discourage frequent track changes and stop edges to minimize conflicts of stopping- and through-traffic.
Priority is set based on inferred track direction based on using \texttt{netconvert} direction priority, and node order \cite{ForwardBackwardLeft}.
Stop edges were extended to accommodate vehicles stopping within one signal block.

\subsection{Stations}
\label{subsec:station_analysis}
First, we extract all relevant station information from OSM, building a structured framework of stop areas, platforms, platform edges, and stop positions. These elements are then combined with the stops from GTFS to build spatial clusters. All elements that are within 500 meters of each other are clustered and assumed to be part of the same station.

Within a station, we attempt to attribute all applicable information derived from platforms, platform edges, and the overall station to the stop positions. Using that information, we construct a penalized mapping from each GTFS stop $h$ to all $m$ simulation stops:
\begin{equation}
    \mathit{StopMap}\!:\; h \to \left\{ (s_{sim,j}, pen_j) \right\}^{m}, \quad 0 < j \leq m.
\end{equation}

The penalty $pen_j$ consists of a component for the confidence in assignment $pen_{ass}$ plus the distance $dist$ from the GTFS stop to the track the stop position is on. This approach allows us to determine preference for track and stop usage without discarding information and disqualifying alternative stop options.

\subsection{Routing}
\label{subsec:routing}

\begin{algorithm}[tb]
    \captionsetup{justification=raggedright,singlelinecheck=false}
    \caption{outerRouter}
    \label{alg:outer_router}
    \begin{algorithmic}
        \State \textbf{Input:} Route $r \gets ((h_1,t_1),(h_2,t_2),\ldots,(h_n,t_n))$
        \State \textbf{Initialize:} $V \gets \emptyset$, $E \gets \emptyset$, $w_s \gets \emptyset$, $w_e \gets \emptyset$
        
        \For{$i \gets 1$ \textbf{to} $n$} \Comment{Build node set and assign penalties}
            \State $S_i \gets \mathit{StopMap}(h_i)$ 
            \For{$(s_{i,j}, p_{i,j}) \in S_i$}
                \State $V_i \gets V_i \cup \{s_{i,j}\}$
                \State $w_s(s_{i,j}) \gets p_{i,j} + \text{additionalPenalties}(s_{i,j}, t_i)$ \Comment{Node weight (penalty)}
            \EndFor
            \State $V = V \cup V_i$
        \EndFor

        \For{$i \gets 1$ \textbf{to} $n-1$} \Comment{Connect subsequent stations}
            \State $S_i \gets \mathit{StopMap}(h_i)$
            \State $S_{i+1} \gets \mathit{StopMap}(h_{i+1})$
            \For{$s_{i,j} \in S_i$}
                \For{$s_{i+1,k} \in S_{i+1}$}
                    \State $E \gets E \cup \{(s_{i,j},s_{i+1,k})\}$
                    \State $w_e((s_{i,j},s_{i+1,k})) \gets \text{innerRoute}(s_{i,j}, s_{i+1,k})$ \Comment{Precomputed Values}
                    \If{$\text{reversing}$}
                        \State $w_e((s_{i,j},s_{i+1,k})) \gets w_e((s_{i,j},s_{i+1,k})) + \text{reversalPenalty}$
                    \EndIf
                \EndFor
            \EndFor
        \EndFor

        \State $g \gets G(V,E,w_s,w_e)$
        \State $path \gets \text{shortestPath}(g,V_1,V_n)$
        \State \Return $path$
    \end{algorithmic}
\end{algorithm}

The routing of trips is implemented via a two-layered, weighted, nested shortest-path approach. This architecture not only bridges the gap between high-level GTFS schedules and microscopic simulation constraints but also ensures high error resistance.
A proper routing algorithm is necessary to avoid simple issues that arise in the SUMO-tools pipeline, such as aggressive routing on one side of parallel tracks (Fig.~\ref{fig:curveRouting}).

Compared to a fixed-stop approach, it handles local data inconsistencies more effectively because it can change multiple parameters simultaneously. 

\subsubsection{Formal Definitions}
We define $S$ as the set of all stops and $T$ as the set of all stop times. Let $R \subseteq \{S \times T\}^*$ be the set of all routes, where the subscripts $_{sim}$ and $_\mathit{gtfs}$ denote the simulation and GTFS domains, respectively. 
A scheduled route $r_\mathit{gtfs} \in R_\mathit{gtfs}$ with $n$ stops is defined as:
\begin{equation}
    r_\mathit{gtfs} = \left((h_1, t_1), (h_2, t_2), \dots, (h_n, t_n)\right)
\end{equation}
where each pair $(h_i, t_i)$ represents a stop identifier and its scheduled time.

\subsubsection{The Two-Layer Approach}
The routing logic is split into an internal micro-router and an external macro-router to balance physical accuracy with network flexibility:

\begin{itemize}
    \item \textbf{Inner Layer (Micro-Routing):} This layer utilizes a \texttt{duarouter} to determine the physical traversal cost between specific platform tracks. It defines the function $\mathit{innerRoute}(u, v)$, which calculates the distance or travel time between subsequent stations. If no physical path exists, it returns $\infty$. These values are computed before running the outer layer.
    \item \textbf{Outer Layer (Macro-Routing):} This layer constructs a directed graph $G=(V, E)$ for each route. The nodes $V$ represent the simulation stops $s_{i,j}$ mapped from GTFS stop $h_i$ via the $\mathit{StopMap}$ function (Alg.~\ref{alg:outer_router}).
\end{itemize}

\begin{figure}[tb]
    \centering
    \footnotesize
    \begin{tikzpicture}
        % Define named nodes
        \node[fill=black, circle, inner sep=2pt, label=left:Start] (Start) at (0,0) {};
        \node[fill=black, circle, inner sep=2pt, label=right:Destination] (End) at (7,0) {};

        % Station 1 bounding box
        \draw[rounded corners] (1,-2) rectangle (2,2);
        \node[above] at (1.5,2) {Station 1};

        % Station 1: in-nodes on left edge, out-nodes on right edge
        \node[fill=black, circle, inner sep=1pt] (P1in1)  at (1,1.6) {};
        \node[fill=black, circle, inner sep=1pt] (P1out1) at (2,1.6) {};
        \node[fill=black, circle, inner sep=1pt] (P1in2)  at (1,0.6) {};
        \node[fill=black, circle, inner sep=1pt] (P1out2) at (2,0.6) {};
        \node[fill=black, circle, inner sep=1pt] (P1in3)  at (1,-0.4) {};
        \node[fill=black, circle, inner sep=1pt] (P1out3) at (2,-0.4) {};
        \node[fill=black, circle, inner sep=1pt] (P1in4)  at (1,-1.4) {};
        \node[fill=black, circle, inner sep=1pt] (P1out4) at (2,-1.4) {};

        % Platform labels centered inside the station box (below the platform level)
        \node[below, font=\scriptsize] at (1.5,1.6) {$P_{1,1}$};
        \node[below, font=\scriptsize] at (1.5,0.6) {$P_{1,2}$};
        \node[below, font=\scriptsize] at (1.5,-0.4) {$P_{1,3}$};
        \node[below, font=\scriptsize] at (1.5,-1.4) {$P_{1,4}$};

        % Internal station edges (in -> out)
        \draw[->] (P1in1) -- node[pos=0.5, fill=white, inner sep=1pt, font=\scriptsize] {$0.15$} (P1out1);
        \draw[->, blue, very thick, dashed] (P1in2) -- node[pos=0.5, fill=white, inner sep=1pt, font=\scriptsize] {$0.1$} (P1out2);
        \draw[->, red, ultra thick, dash dot] (P1in3) -- node[pos=0.5, fill=white, inner sep=1pt, font=\scriptsize] {$0.15$} (P1out3);
        \draw[->, orange, very thick, densely dotted] (P1in4) -- node[pos=0.5, fill=white, inner sep=1pt, font=\scriptsize] {$0.5$} (P1out4);

        % Start connects to Station 1 in-nodes
        \draw[->] (Start) to[bend left=30] (P1in1);
        \draw[->, blue, very thick, dashed] (Start) to[bend left=10] (P1in2);
        \draw[->, red, ultra thick, dash dot] (Start) to[bend right=10] (P1in3);
        \draw[->, orange, very thick, densely dotted] (Start) to[bend right=30] (P1in4);

        % Station 2 bounding box
        \draw[rounded corners] (3,-2) rectangle (4,2);
        \node[above] at (3.5,2) {Station 2};

        % Station 2: in-nodes on left edge, out-nodes on right edge
        \node[fill=black, circle, inner sep=1pt] (P2in1)  at (3,1.6) {};
        \node[fill=black, circle, inner sep=1pt] (P2out1) at (4,1.6) {};
        \node[fill=black, circle, inner sep=1pt] (P2in2)  at (3,0.6) {};
        \node[fill=black, circle, inner sep=1pt] (P2out2) at (4,0.6) {};
        \node[fill=black, circle, inner sep=1pt] (P2in3)  at (3,-0.4) {};
        \node[fill=black, circle, inner sep=1pt] (P2out3) at (4,-0.4) {};
        \node[fill=black, circle, inner sep=1pt] (P2in4)  at (3,-1.4) {};
        \node[fill=black, circle, inner sep=1pt] (P2out4) at (4,-1.4) {};

        % Platform labels centered inside the station box (below the platform level)
        \node[below, font=\scriptsize] at (3.5,1.6) {$P_{2,1}$};
        \node[below, font=\scriptsize] at (3.5,0.6) {$P_{2,2}$};
        \node[below, font=\scriptsize] at (3.5,-0.4) {$P_{2,3}$};
        \node[below, font=\scriptsize] at (3.5,-1.4) {$P_{2,4}$};

        % Internal station edges (in -> out)
        \draw[->] (P2in1) -- node[pos=0.5, fill=white, inner sep=1pt, font=\scriptsize] {$0.5$} (P2out1);
        \draw[->] (P2in2) -- node[pos=0.5, fill=white, inner sep=1pt, font=\scriptsize] {$0.4$} (P2out2);
        \draw[->, red, ultra thick, dash dot] (P2in3) -- node[pos=0.5, fill=white, inner sep=1pt, font=\scriptsize] {$0.1$} (P2out3);
        \draw[->, orange, very thick, densely dotted] (P2in4) -- node[pos=0.5, fill=white, inner sep=1pt, font=\scriptsize] {$0.3$} (P2out4);

        % Inter-station edges: Station 1 out-nodes -> Station 2 in-nodes
        \draw[->] (P1out1) -- node[pos=0.50, fill=white, inner sep=1pt] {$1$} (P2in1);
        \draw[->] (P1out1) -- node[pos=0.70, fill=white, inner sep=1pt] {$2$} (P2in2);

        \draw[->] (P1out2) -- node[pos=0.30, fill=white, inner sep=2pt] {$2$} (P2in1);
        \draw[->] (P1out2) -- node[pos=0.50, fill=white, inner sep=1pt] {$2$} (P2in2);

        \draw[->, blue, very thick, dashed] (P1out2) -- node[pos=0.50, fill=white, inner sep=1pt] {$6$} (P2in3);

        \draw[->, red, ultra thick, dash dot] (P1out3) -- node[pos=0.50, fill=white, inner sep=1pt] {$1$} (P2in3);
        \draw[->] (P1out3) -- node[pos=0.70, fill=white, inner sep=1pt] {$2$} (P2in4);

        \draw[->] (P1out4) -- node[pos=0.30, fill=white, inner sep=1pt] {$2$} (P2in3);
        \draw[->, orange, very thick, densely dotted] (P1out4) -- node[pos=0.50, fill=white, inner sep=1pt] {$1.5$} (P2in4);

        % Station 3 bounding box
        \draw[rounded corners] (5,-2) rectangle (6,2);
        \node[above] at (5.5,2) {Station 3};

        % Station 3: in-nodes on left edge, out-nodes on right edge
        \node[fill=black, circle, inner sep=1pt] (P3in1)  at (5,1.6) {};
        \node[fill=black, circle, inner sep=1pt] (P3out1) at (6,1.6) {};
        \node[fill=black, circle, inner sep=1pt] (P3in2)  at (5,0.6) {};
        \node[fill=black, circle, inner sep=1pt] (P3out2) at (6,0.6) {};
        \node[fill=black, circle, inner sep=1pt] (P3in3)  at (5,-0.4) {};
        \node[fill=black, circle, inner sep=1pt] (P3out3) at (6,-0.4) {};
        \node[fill=black, circle, inner sep=1pt] (P3in4)  at (5,-1.4) {};
        \node[fill=black, circle, inner sep=1pt] (P3out4) at (6,-1.4) {};

        % Platform labels centered inside the station box (below the platform level)
        \node[below, font=\scriptsize] at (5.5,1.6) {$P_{3,1}$};
        \node[below, font=\scriptsize] at (5.5,0.6) {$P_{3,2}$};
        \node[below, font=\scriptsize] at (5.5,-0.4) {$P_{3,3}$};
        \node[below, font=\scriptsize] at (5.5,-1.4) {$P_{3,4}$};

        % Internal station edges (in -> out)
        \draw[->] (P3in1) -- node[pos=0.5, fill=white, inner sep=1pt, font=\scriptsize] {$0.4$} (P3out1);
        \draw[->] (P3in2) -- node[pos=0.5, fill=white, inner sep=1pt, font=\scriptsize] {$0.35$} (P3out2);
        \draw[->, blue, very thick, dashed] (P3in3) -- node[pos=0.5, fill=white, inner sep=1pt, font=\scriptsize] {$0.2$} (P3out3);
        \draw[->, red, ultra thick, dash dot] (P3in4) -- node[pos=0.5, fill=white, inner sep=1pt, font=\scriptsize] {$0.3$} (P3out4);

        % Inter-station edges: Station 2 out-nodes -> Station 3 in-nodes
        \draw[->] (P2out1) -- node[pos=0.50, fill=white, inner sep=1pt] {$2$} (P3in1);
        \draw[->] (P2out1) -- node[pos=0.70, fill=white, inner sep=1pt] {$3$} (P3in2);

        \draw[->] (P2out2) -- node[pos=0.30, fill=white, inner sep=2pt] {$3$} (P3in1);
        \draw[->] (P2out2) -- node[pos=0.50, fill=white, inner sep=1pt] {$1$} (P3in2);

        \draw[->, blue, very thick, dashed] (P2out3) -- node[pos=0.50, fill=white, inner sep=1pt] {$7$} (P3in3);
        \draw[->, red, ultra thick, dash dot] (P2out3) -- node[pos=0.70, fill=white, inner sep=1pt] {$2$} (P3in4);

        \draw[->] (P2out4) -- node[pos=0.30, fill=white, inner sep=1pt] {$2$} (P3in3);
        \draw[->, orange, very thick, densely dotted] (P2out4) -- node[pos=0.50, fill=white, inner sep=1pt] {$1$} (P3in4);

        % Station 3 out-nodes connect to End (bent)
        \draw[->] (P3out1) to[bend left=25] (End);
        \draw[->] (P3out2) to[bend left=10] (End);
        \draw[->, blue, very thick, dashed] (P3out3) to[bend right=10] (End);
        \draw[->, red, ultra thick, dash dot] (P3out4) to[bend right=25] (End);
    \end{tikzpicture}
    \caption{Example routing showing a fixed stop routing (blue, dashed), a lowest edge cost traversal (orange, dotted), and routing with both edge and node weights (red, dash--dotted). The top two rows represent tracks with a preferred direction left, and the bottom two rows have a preferred direction right.}
    \label{fig:routing}
\end{figure}
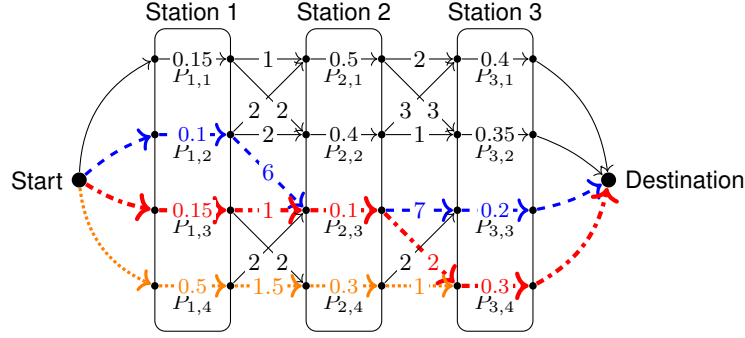
\begin{figure}[tb]
    \centering
    \includegraphics[width=0.543\linewidth]{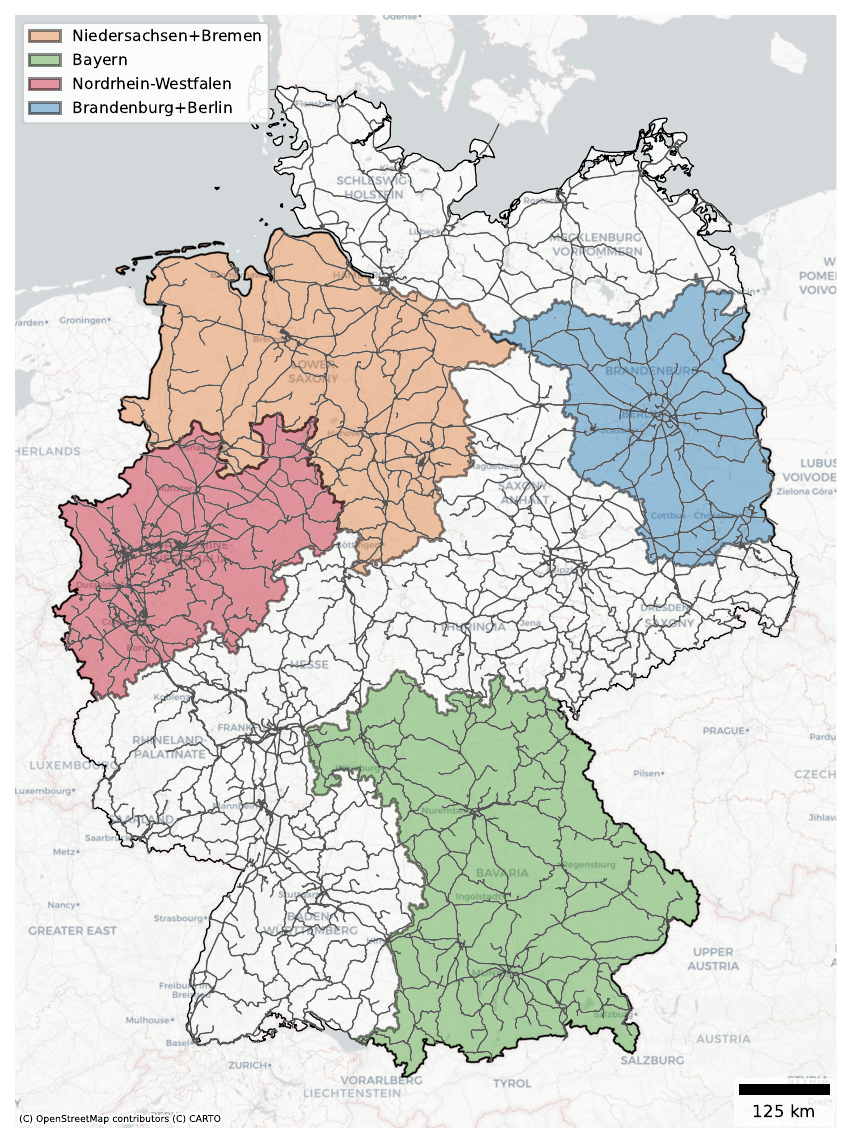}
    \caption{Study areas for this paper. We applied the GROSS pipeline on subregions of Germany to compare its performance using only SUMO tools.}
    \label{fig:map}
\end{figure}

Edge weights $w_e$ are derived from the inner router's cost, while node weights $w_s$ represent platform-specific penalties. This redundancy provides error resistance: if inconsistent, wrong, or missing data breaks a connection, the graph is still able to select the holistically best solution (Fig.~\ref{fig:routing}).

\subsubsection{Node Penalty Calculation}
The penalty $p_{i,j}$ assigned to a simulation stop $s_{sim,i,j}$ is designed to minimize platform conflict through stochastic and temporal weighting. The penalty is formulated as the sum of three distinct factors:
\begin{equation}
    \text{additionalPenalties}(s_{i,j}, t_i) = R + U_{short}(t_i) + U_{long}(t_i)
\end{equation}
where $R$ is a random variable sampled from the uniform distribution $\mathcal{U}(0, 8)$. The temporal components, $U_{short}$ and $U_{long}$, are defined by linear ramps based on the planned arrival $t_{arr}$ and departure $t_{dep}$. $U_{short}$ provides a peak of $64$ during the planned stop (ramping up over 5 minutes and down over 10 minutes). In contrast, $U_{long}$ provides a broader baseline penalty of $16$, starting 1 hour before arrival and persisting until 2 hours after departure to account for extended platform occupancy.

\subsubsection{Route Repair}
\label{subsubsec:route_repair}

The route repair applies in addition to the flexible routing some adaptations based on issues we have found in the data and processing with 18\% of trips affected (Tab. \ref{tab:repair}).
For routes that stop at a station with a single stop platform that, at the planned time, is already occupied, we allow computing the route to the following station.
For missing stops (due to stops incorrectly tagged in OSM or outside the simulated networks) or sections that would require an impossible speed to traverse within the scheduled time, we allow skipping the stop by removing the vehicle at the previous stop and inserting it at the following stop after the scheduled travel time (while retaining delays) using the \texttt{jump} stop attribute in SUMO. 

\subsection{Simulation Experiments}
\label{subsec:sim_experiments}

Using SUMO (v1.26), we created and ran multiple scenarios with the GROSS pipeline on sub-regions of Germany.
We sampled the GTFS dataset to create scenarios representing Wednesdays and Fridays between 14--28 January 2026, a period without public or school holidays.
We selected Wednesdays as a representative weekday and Fridays to capture additional long-distance routes connecting Germany to other European cities.
To highlight the impact of teleportation effects (and the resulting fix), we set \texttt{time-to-teleport} to 1800~s and \texttt{time-to-teleport.deadlocks} to 300~s.
For each scenario, we ran 10 replicates, each with a different seed.
We focused on four sub-regions to compare the GROSS pipeline against a SUMO tool pipeline.
We focus on the sub-regions of Bayern, Brandenburg (+\,Berlin), Niedersachsen (+\,Bremen), and Nordrhein-Westfalen to compare the GROSS pipeline against a SUMO tool pipeline (Fig.~\ref{fig:map}).

\section{Results}

Our processing pipeline enables substantially faster generation of usable SUMO rail scenarios than the vanilla SUMO toolchain (Tab. \ref{tab:process_time}).
Across our four evaluation regions (Brandenburg (+ Berlin), Niedersachsen (+ Bremen), Bayern, and Nordrhein-Westfalen), the baseline SUMO workflow requires 121~minutes (Niedersachsen) to 1638~minutes (Bayern) to generate route files alone.
In contrast, GROSS completes the route generation in 12~minutes (Brandenburg) and 49~minutes (Nordrhein-Westfalen).
At national scale, this gap widens further: GROSS generates a Germany-wide routing in 190~minutes, whereas the baseline \texttt{gtfs2pt} route generation took 106~hours.

\begin{table}[tbp]
  \centering
  \caption{Teleports per vehicle and vehicle count by region and scenario (averaged across seeds and dates). Teleportations are reduced by a factor of 1.7 to 76.8$\times$ compared to SUMO. }
  \label{tab:teleports}
  \begin{tabular}{lrrrr}
    \toprule
    & \multicolumn{2}{c}{Trips} & \multicolumn{2}{c}{Teleports/veh} \\ 
    Region & SUMO & GROSS & SUMO & GROSS \\ 
    \midrule
Bayern & 7451 & 7341 & 1.015 & 0.583 \\
Brandenburg & 2361 & 2381 & 0.736 & 0.045 \\
Niedersachsen & 3165 & 2976 & 0.764 & 0.017 \\
Nordrhein-Westfalen & 6295 & 6199 & 0.922 & 0.012 \\
\midrule 
Germany & 37710 & 35925 & 0.991 & 0.219 \\
    \bottomrule
  \end{tabular}
\end{table}

\subsection{German sub-regions}
\begin{figure}[tb]
    \centering
    \includegraphics[width=0.9\linewidth]{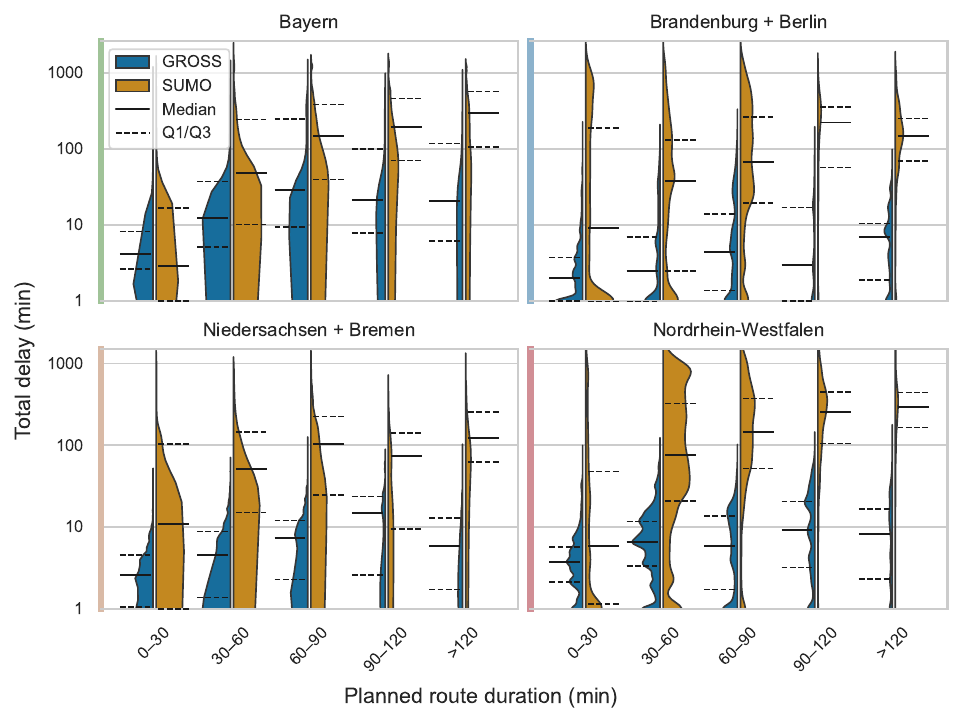}
    \caption{Regional-level comparison of vehicle delay distributions, grouped by planned route duration.
    The ticks show the median and 1st and 3rd quartiles of each distribution.
    % A zoomed version is provided for Nordrhein-Westfalen to show the performance of GROSS across the planned route durations.
    Across all regions, GROSS yields a markedly smaller delay range. Rare outliers ($<$5\% of trips) still occur, with the largest observed delays in Bayern.}
    \label{fig:delays_regions}
\end{figure}
A central challenge when scaling the vanilla SUMO rail pipeline is the prevalence of teleportation artifacts.
To prevent extreme queueing and gridlock, SUMO can teleport vehicles that have been blocked at a junction for longer than a fixed threshold.
While this mechanism introduces only minor distortions in road-traffic settings, it is highly unrealistic for rail operations because it breaks vehicle continuity and effectively bypasses capacity constraints.
Thus we deliberately evaluate our scenarios with teleportation thresholds of 30 minutes.

Even with a conservative teleportation threshold of 30~minutes, the baseline scenarios exhibit frequent teleports (Tab.~\ref{tab:teleports}), which lead to extreme schedule deviations.
In particular, the median delay in the baseline reaches 42--88~minutes (Fig.~\ref{fig:delays_regions}), reflecting compounding effects of interactions, missed meets/overtakes, and repeated teleport triggers.
Nordrhein-Westfalen suffers from the worst delays, with half of the trips delayed by more than 88~minutes.

GROSS reduces teleports by a factor of 1.7--76.8$\times$ relative to SUMO and correspondingly shifts the delay distributions downward across all planned-duration bins (for spatial distribution, see Fig.~\ref{fig:teleport_map}).
As expected, longer trips accumulate more variability because they traverse more junctions and are exposed to more opportunities for interaction-induced delay.
However, in GROSS these delays remain bounded for the majority of trips (median of 2.5--10~minutes); only a small fraction of services (below 5\%) exhibits large tail delays, and these are substantially less severe than in the baseline.

The teleportation reduction compared to the baseline is primarily driven by GROSS route processing, which 
(i) reconciles timetable stop sequences with the rail topology, 
(ii) enforces consistent track usage at stations and along corridor segments to avoid spurious conflicts or deadlocks, and 
(iii) avoids large detours due to infeasible itineraries (e.g. going from $A\to C\to B$ where $C$ is an unnecessary non-stopping detour) introduced by noisy schedule-to-network matching.
Together, these steps reduce artificial bottlenecks (e.g., multiple trains being forced through an incorrect edge or platform) and produce smoother headway patterns, lowering the probability that trains remain blocked long enough to trigger teleportation.

\begin{figure}[tb]
    \centering
    \includegraphics[]{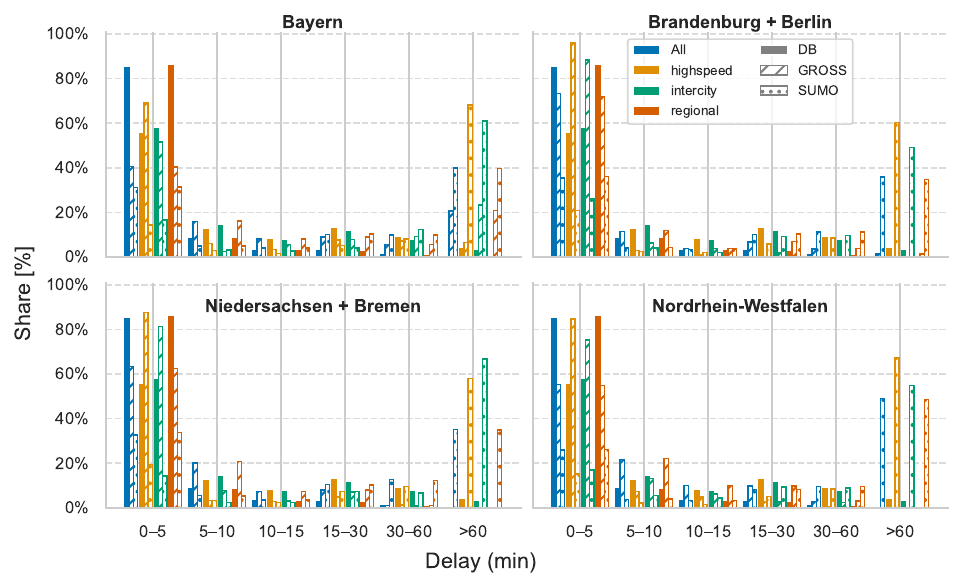}
    \caption{Share of vehicle-stops over different delay durations. Compared to station-level delay statistics from Deutsche Bahn (DB) at the national level, the GROSS scenario reproduces the overall shape of the distribution but still exhibits an over-representation of large delays. In contrast, the SUMO pipeline scenario produces a large amount of trains ($>$30\%) that are delayed more than 60 minutes across the four regions.  }
    \label{fig:cities_gross_vs_db}
\end{figure}

To contextualize the realism of these synthetic scenario, we compare simulated stop delays against real-world delay distributions reported by Deutsche Bahn (DB), the German national railway operator.
We obtain station-level delay statistics for the period December 2025--January 2026 using the data collection scripts by \url{https://github.com/piebro/deutsche-bahn-data}. % Ref.~\cite{brommelPiebroDeutschebahndata2026}. %
For the simulated scenario, we compute per-stop delay as the difference between planned and realized arrival times at each stop, and aggregate delays into the same bins used by the DB statistics.

With this, we are now able to compare the performance of SUMO and GROSS side-by-side in context with the train statistics from DB.
In Fig.~\ref{fig:cities_gross_vs_db}, we see that SUMO suffers from a lot of routes having more than 60\,minutes delays.
For the region of Bayern, we see that, although it performs better than the SUMO pipeline, our GROSS pipeline has some trouble with the regional train routes, with around 21\% having more than 60\,minutes delays. We suspect that this is due to the station topology of München Hbf.
However, apart from Bayern, the other regions perform much better than the SUMO-generated scenarios, albeit still below the real-world statistics of DB.

\subsection{Germany Train Scenario}
\begin{figure}[tb]
    \centering
    \includegraphics[width=\linewidth]{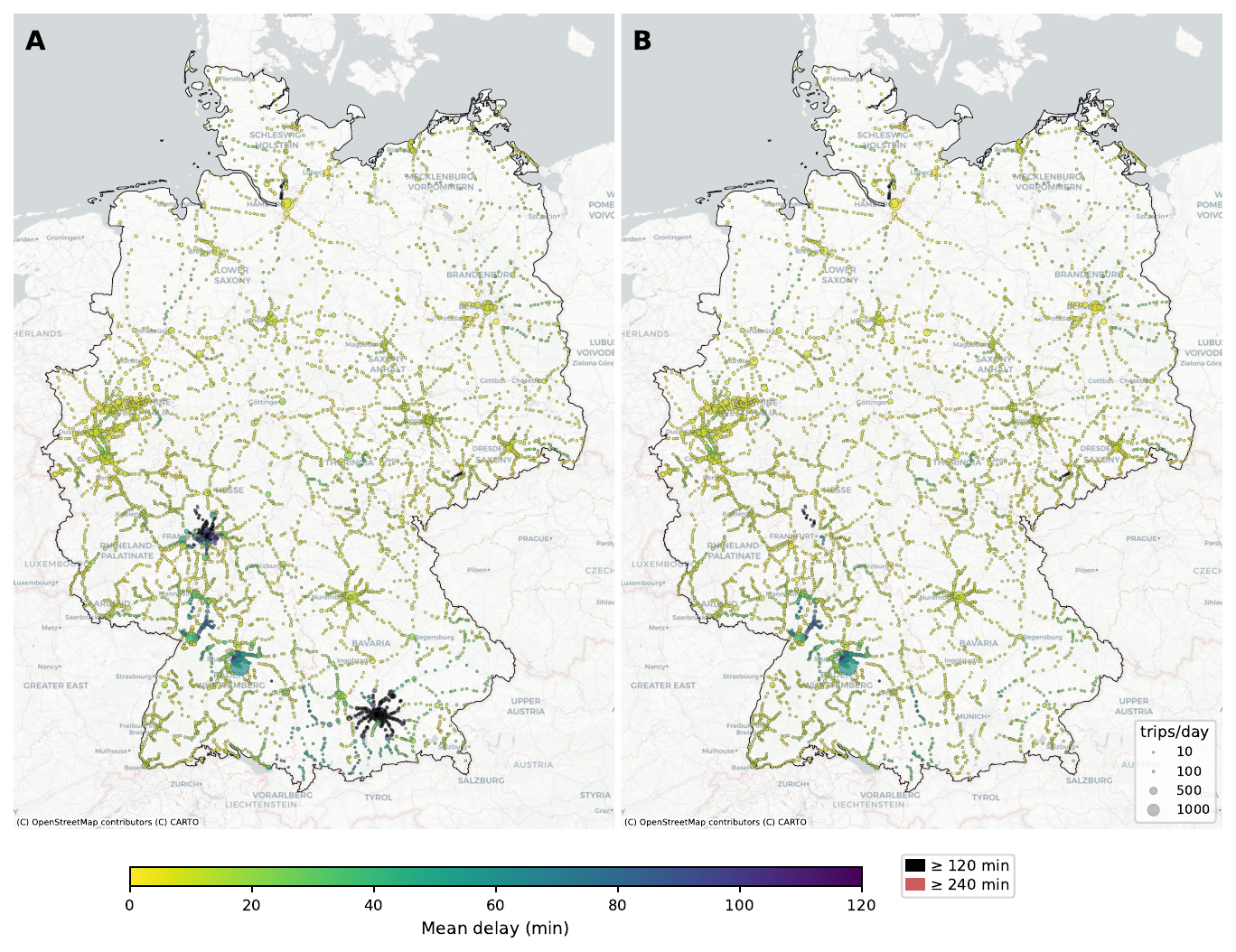}
    \caption{Mean delay (in minutes) observed at each station cluster for Germany. The sizes of the bubbles also show the number of train routes per day passing through the station cluster. (A) All trips included in analysis. (B) Trips with stops passing through München and Frankfurt am Main filtered out.}
    \label{fig:gross_vs_db_map}
\end{figure}

The performance and robustness improvements of GROSS enable us to generate and simulate a Germany-wide rail scenario in SUMO.
After running the GROSS pipeline for Germany, we observe that stations in certain regions have disproportionately large delays.
These can be seen in the cities of München and Frankfurt am Main (Fig.~\ref{fig:gross_vs_db_map}A).

We hypothesize that some of these large delays are linked to challenging local layouts, such as terminating stations (where trains cannot pass through) with a limited through-track capacity. 
We suspect that this happens as trips get routed via through-tracks that would normally stop at a terminal platform due to shorter connecting tracks.

\begin{figure}[tb]
    \centering
    \includegraphics[width=\linewidth]{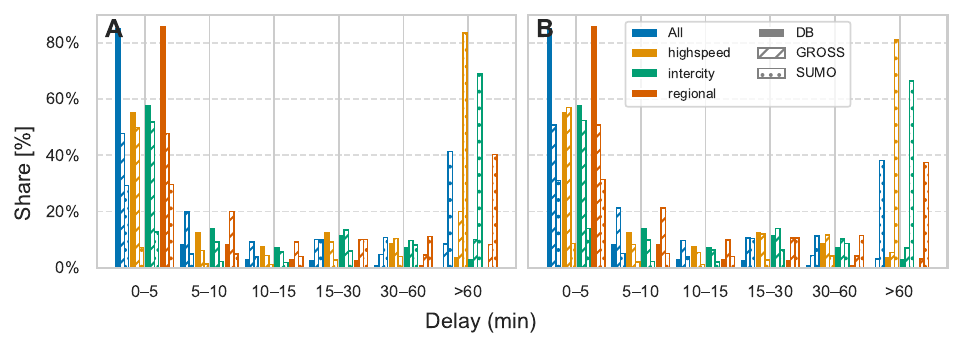}
    \caption{Share of vehicle-stops over different delay durations. We show the stop-level statistics for (A) all trips included in analysis, and (B) trips with stops passing through München and Frankfurt am Main filtered out. Filtering out the trips affects mostly the trips with delays exceeding 60 minutes.}
    \label{fig:gross_vs_db}
\end{figure}

Instead, we try to see whether this is an isolated effect by excluding all trips that pass through these cities from our analysis.
That is, we still simulated Germany with all trips, but filter out this subset of trips (around 15\% of stop events) from the analysis (Fig.~\ref{fig:gross_vs_db_map}B).
Although we show that GROSS substantially reduces unrealistically extreme delay behavior compared to the vanilla SUMO pipeline, Fig.~\ref{fig:gross_vs_db} nevertheless shows a delay distribution still produces too many high-delay events compared to the DB profile.
In our analysis, these tail delays are often associated with local deadlocks (e.g., opposing trains meeting on a single-track segment without a feasible resolve) and can cascade to downstream services when headways are tight.
Addressing this likely requires more detailed modeling of operational rules and infrastructure constraints (e.g., passing-loop usage, dispatching/priority policies, and single-track conflict resolution), which we leave for future work.

\section{Discussion}

GROSS tackles two recurring practical obstacles when building large-scale rail scenarios in SUMO: long preprocessing times and simulations that become unstable due to routing errors and the resulting teleportation artifacts. By combining topology-aware route construction with systematic validation and targeted repair, the pipeline produces scenarios that are faster to generate and operationally more plausible. This is reflected in substantially reduced teleport rates and a tighter delay distribution, suggesting that a large fraction of previously observed ``instability'' was caused by inconsistent inputs and brittle routing rather than by unavoidable effects of scale.

The results also clarify what GROSS does and does not solve. The pipeline improves the consistency between the timetable and the network representation, but it cannot compensate for missing information in the underlying data sources. GTFS feeds are primarily meant for passenger-facing schedule dissemination and often omit infrastructure- and operations-relevant detail such as platform-track assignments, passing loop usage, rolling-stock constraints, or operator-specific dispatching rules. Consequently, even after cleaning and repair, the simulated system is best interpreted as an operationally feasible \,but simplified\, realization of the published service plan.

Several remaining failure modes are also tied to limitations of the SUMO rail model and the configuration used in this work. In particular, the current setup does not fully capture the richness of railway operations, including explicit timetabled meets on single-track, priority- and rule-based dispatching, and recovery margins that buffer small perturbations. These modeling gaps can manifest as unrealistic deadlocks or exaggerated tail delays in specific network regions. The most prominent issues observed in our experiments were:
\begin{itemize}
    \item \textbf{Coupling and splitting:} coupled trains are represented as independent vehicles, which can amplify delays when they traverse long blocks and compete for scarce capacity.
    \item \textbf{Single-track conflict resolution:} on single-track sections with multiple uncoordinated signal junctions, local look-ahead can admit trains from both ends and trigger deadlocks.
    \item \textbf{Platform assignment and re-planning:} trains are not re-routed online to alternative platforms, so temporary platform conflicts can cascade into system-wide delay propagation.
    \item \textbf{Residual schedule noise:} mistagged or malformed GTFS trips may still be imported and then routed in implausible ways (e.g., repeated reversals), requiring detection and filtering via exception lists.
\end{itemize}

Despite these limitations, the evaluation demonstrates that the approach scales from subregional deployments to a nation-scale scenario without changing the conceptual workflow. Beyond Germany, GROSS is directly applicable wherever (i) a rail network can be extracted (e.g., from OpenStreetMap) and (ii) service schedules are available in GTFS or can be converted to it. The GTFS standard is widely available from different agencies or at sites like the Mobility Database \cite{noauthor_mobility_nodate}. 
Alternatively, EU members are mandated to provide a national access point for NeTeX feeds \cite{noauthor_national_nodate} that can be converted to GTFS using external tools \cite{noauthor_enturnetex-gtfs-converter-java_2026}.

Although we focused on a rail-only setup, the generated rail network is also intentionally composable with existing SUMO scenarios: it can be merged with city- or region-scale road and active-mobility networks via \texttt{netconvert} (using the arguments \texttt{-s train.net.xml,notrain.net.xml} to merge networks).
This means that GROSS can complement other preexisting, more car- and bus-focused scenarios for building multi-modal simulations in which rail interacts with buses, private vehicles, and pedestrians, thereby lowering the barrier to constructing more holistic digital twins and enabling comparative studies under consistent modeling assumptions.

% If again we would do it without splitting
% No tagging along the distinction we attempted to split by
In this work we found that there is no consistent tagging in regards to how we distinguished light-rail and heavy-rail as heavy-rail being all systems that are interoperable with the national network.
Thus we see advantages in simulating all rail traffic (tram, light-rail, heavy-rail, narrow-gauge) to reduce vulnerability to tagging inconsistencies and simplify combined systems (e.g. S-Bahn Karlsruhe operating on the tram, and rail network).

Looking forward, two directions appear most important for closing the remaining realism gap. First, robust station/stop matching across heterogeneous identifiers (e.g., GTFS stop IDs, operator station codes, and external delay statistics) would benefit from standardized mappings and automated alignment routines. Second, reducing the high-delay tail will likely require incorporating additional operational constraints, dispatching priorities, and control logic, especially for single-track sections.

\section{Conclusion and Future Work}
We introduced GROSS, a preprocessing pipeline for generating large-scale SUMO rail scenarios that is both faster than the vanilla SUMO toolchain and substantially more stable in simulation.
Across multiple German regions, GROSS reduces teleportation events and produces significantly lower and less variable delay distributions.
At a national scale, the pipeline enables end-to-end generation of a Germany-wide scenario and supports comparison against operator-reported delay statistics while highlighting remaining gaps in the high-delay tail.

Future work will focus on (i) improving data alignment and coverage (e.g., robust station/stop matching and richer infrastructure metadata when available), (ii) improving operational realism through enhanced dispatching and single-track conflict handling, and (iii) tighter integration with multi-modal, city-scale digital twins by merging GROSS rail networks with existing SUMO road and public-transport scenarios using \texttt{netconvert}.
Together, these directions move toward scalable, transferable, and multi-modal simulation environments that can better support urban planning and policy analysis across diverse geographies.

\section*{Data availability statement}
All code used to generate the resulting scenarios in the work will be made available at \url{https://github.com/ethz-coss/GROSS}.

\section*{Author contributions}
J.P.: Conceptualization; methodology; software; data curation; formal analysis; validation; visualization; investigation; writing—original draft; writing—review and editing.

\noindent D. D.: Conceptualization; supervision; methodology; software; formal analysis; validation; visualization; investigation; writing—original draft; writing—review and editing.

\section*{Competing interests}
The authors declare that they have no competing interests.

\section*{Acknowledgements}
J.P. and D.D. would like to acknowledge Prof. Dirk Helbing for his support and insightful feedback.

\printbibliography[heading=references]

\clearpage

\appendix
\section*{Appendix}

% Appendix figure/table numbering like A1, A2, ... (keep appendix sections unnumbered)
\setcounter{figure}{0}
\setcounter{table}{0}
\renewcommand\thefigure{A\arabic{figure}}
\renewcommand\thetable{A\arabic{table}}

\subsection*{Teleportation heatmap}
\begin{figure}[h]
    \centering
    \includegraphics[width=\linewidth]{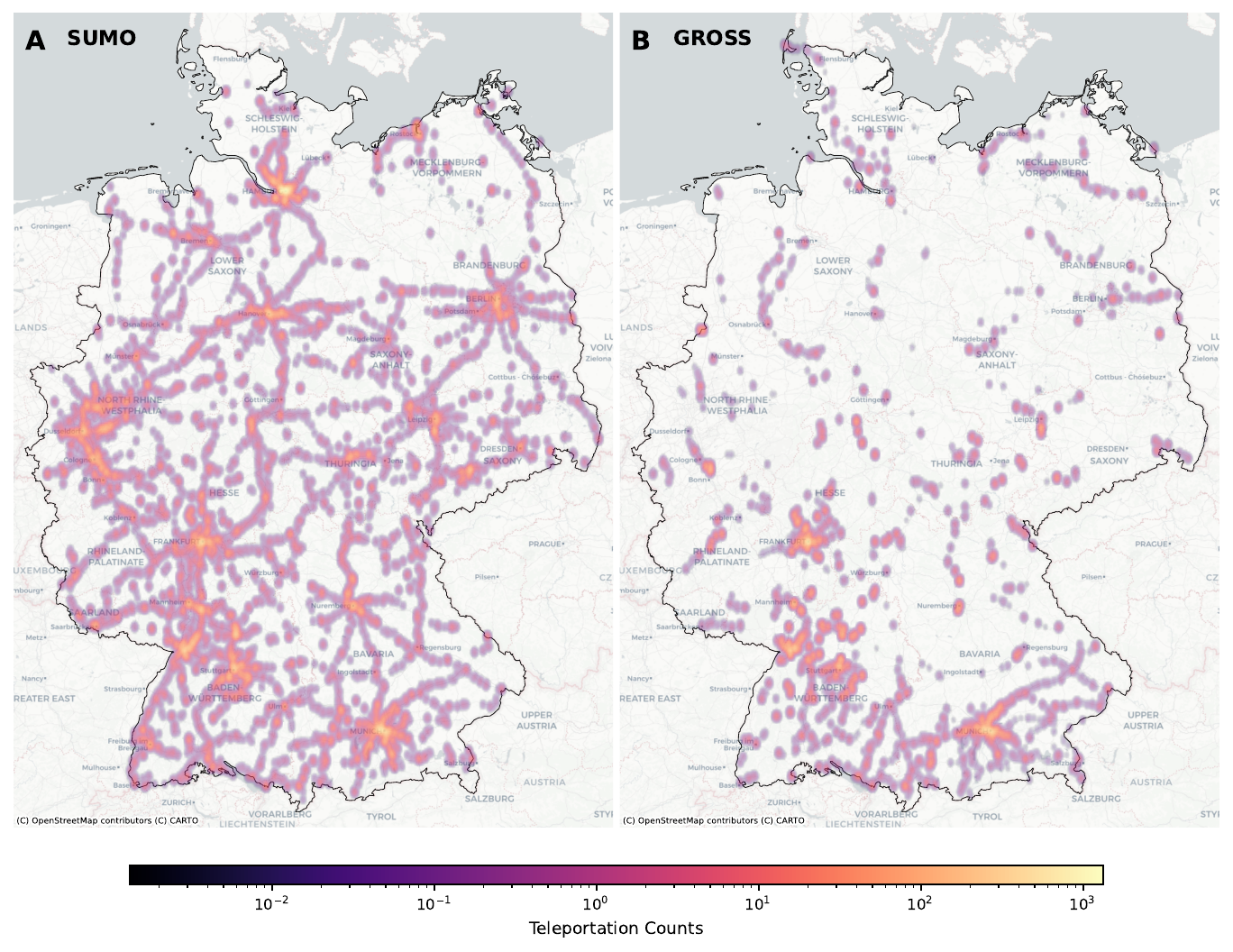}
    \caption{Spatial distributions of teleportation between SUMO (left) and GROSS (right) across Germany. There are far fewer teleportations for GROSS except for the regions of Karlsruhe, München, and Frankfurt am Main.}
    \label{fig:teleport_map}
\end{figure}

\subsection*{Processing Time Table}
\begin{table}[H]
    \centering
    \begin{tabularx}{0.6\linewidth}{l >{\raggedleft\arraybackslash}X >{\raggedleft\arraybackslash}X >{\raggedleft\arraybackslash}X}
        \toprule
         & \multicolumn{3}{c}{Processing Time (in minutes)} \\
         Region & \multicolumn{1}{r}{SUMO} & \multicolumn{2}{c}{GROSS} \\
          & \multicolumn{1}{r}{Per Day} & \multicolumn{1}{r}{Base} & \multicolumn{1}{r}{Per Day} \\
        \midrule
        Bayern & 1638 & 30 & 2 \\
        Brandenburg & 222 & 8 & 1 \\
        Niedersachsen & 121 & 15 & 1 \\
        Nordrhein-Westfalen & 941 & 36 & 2 \\
        \midrule
        Germany & 6360 & 369 & 10 \\
        \bottomrule
    \end{tabularx}
    \caption{Processing Time to generate a day of rail traffic. For GROSS split into base cost and additional cost per simulation day.}
    \label{tab:process_time}
\end{table}

\newpage

\subsection*{Repair Table}
\begin{table}[h]
    \centering
    % Percent columns use siunitx's S column type.
    \begin{tabular}{lSSS}
    \toprule
        & \multicolumn{1}{c}{SUMO (\%)} & \multicolumn{2}{c}{GROSS (\%)} \\
        \cmidrule(lr){2-2} \cmidrule(lr){3-4}
        Repair Fix & {Trips} & {Stops} & {Trips} \\
        \midrule
        Single-Stop  & {\textemdash} & 0.038 & 3.2 \\
        Missing Stop & {\textemdash} & 3.1 & 16 \\
        Speed  & 1.4 & {\textemdash} & 0.0 \\
        \bottomrule
    \end{tabular}
    \caption{Number of stops and trips affected by route repair (averaged across seeds and dates in Germany). Some routes are excluded due to manual review.}
    \label{tab:repair}
\end{table}

\end{document}